\documentclass[prd,aps,twocolumn]{revtex4}
\usepackage{graphicx}
\usepackage{color}
\usepackage{xtab}

\begin{document}

\title{ Neutrinos in IceCube from AGN's}
\author{Oleg Kalashev$^1$}
\author{Dmitri  Semikoz$^{2}$}
\author{Igor Tkachev$^1$}
\affiliation{$^{1}$Institute for Nuclear Research of the Russian Academy of Sciences, Moscow  117312, Russia}
\affiliation{$^2$Laboratory of AstroParticle and Cosmology (APC), Paris, France}

\begin{abstract}
Recently IceCube collaboration has reported first evidence for the astrophysical neutrinos. Observation corresponds to the total astrophysical neutrino flux of the order of $3 \cdot 10^{-8}$ \mbox{GeV cm}$^{-2}$s$^{-1}$sr$^{-1}$ in a PeV energy range~\cite{Icecube3years}. Active Galactic Nuclei (AGN) are natural candidate sources for such neutrinos. To model the neutrino creation in AGNs we study photopion production processes on the radiation field of the Shakura-Sunyaev accretion disks in the black hole vicinity. We show that this model can explain detected neutrino flux and avoids, at the same time, existing constraints from the gamma-ray and cosmic ray observations. 
\end{abstract}
\maketitle

\section{Introduction}

Detection of astrophysical neutrinos by the IceCube collaboration \cite{Icecube3years} has opened new era in the high energy astrophysics. Reported excess of neutrinos at $E>30$ TeV energies can be described by a power law $1/E^\alpha$ with $\alpha=2.3 \pm 0.3$, and corresponds to the flux $3 \cdot 10^{-8}$ \mbox{GeV cm}$^{-2}$ s$^{-1}$ sr$^{-1}$ for the sum of three flavors, possibly with cutoff at 3 PeV~\cite{Icecube3years}.  This observation has high significance of 5.7 $\sigma$ and calls for theoretical modeling and explanation.

There are three main  production mechanisms  of high energy neutrinos.  First, galactic cosmic rays produce neutrinos in the proton-proton (proton-nuclei) collisions in the interstellar gas in the disk of our Milky Way Galaxy. Such neutrinos would have energies from sub-GeV up to PeV, but can come only from directions close to the Galactic plane. Interestingly, 3-year IceCube data do show some access in the direction of Galactic plane with  2\% chance probability~\cite{Icecube3years} possibly exhibiting small scale anisotropy near Galactic center. Both signatures can be explained by the neutrino production in the galactic cosmic ray interactions with the interstellar gas. In Ref.~\cite{Joshi:2013aua} it was shown that only at most 0.1 of the observed neutrino events in IceCube can be described by cosmic ray interactions with matter inside Milky Way assuming local density of gas.
However, expected signal is dominated by the flux from spiral arms and/or  Galactic Bar, where Supernova explosion rates, magnetic fields and density of interstellar gas are all much higher than those in the vicinity of the Sun~\cite{NS2013}. Moreover, neutrino flux detected by the IceCube  is consistent~\cite{NS2013} with the power law extrapolation of $E>100$ GeV diffuse gamma-ray flux from the Galactic Ridge, as observed by Fermi telescope, which suggests common origin. As result, 
contribution of Galaxy to neutrino flux can be much higher than 10 \%.

Second, Ultra-High Energy Cosmic Rays (UHECR) interact with intergalactic radiation
and produce secondary EeV neutrinos in pion decays. The latter are called cosmogenic neutrinos  and has been extensively studied theoretically since 1969~\cite{BZ1969} onwards (see e.g.~\cite{reviewGZKneutrinos,Semikoz:2003wv} and references therein).  Expected flux of cosmogenic neutrinos is somewhat model dependent, but even optimistic estimates are at least two orders of magnitude below IceCube signal at PeV energies.  Thus cosmogenic neutrinos are irrelevant in this energy range.

Finally, high energy neutrinos in a wide range of energies, from TeV to 10 PeV, can be produced in a variety of astrophysical sources in decays of charged pions  created in the  proton-photon or proton-proton collisions {\it in situ}.  Various kinds of astrophysical sources of high energy neutrinos were considered prior to the IceCube observation, including AGN's \cite{eichler,stecker,atoyan,halzen,NS2002,Kachelriess:2008qx}, Gamma-Ray  Birst's  \cite{grb}, Star Burst Galaxies \cite{SBG}.

After  IceCube observation the interest to the problem has grown substantially. In a number of recent works~\cite{Cholis:2012kq,Stecker:2013fxa,Murase:2013ffa,Anchordoqui:2013dnh,Dermer:2014vaa,Murase:2014foa} the attempt was made to explain IceCube events by various astrophysical sources of high energy neutrino.

In this paper we are developing model originally proposed in Ref.~\cite{stecker}, where neutrinos arise in interactions of high energy cosmic rays 
accelerated in AGN with photons from the big blue bump. As compared to previous papers developing this concept, we  are attempting to explain the IceCube observation using photopion production by cosmic rays on anisotropic radiation field produced by the realistic Shakura-Sunyaev model of accretion disks~\cite{Shakura}. 

The paper is organized as follows.
In Section~\ref{sec:SS} we present theoretical details of our calculation,  reviewing also the  observation knowledge about black hole accretion discs and their radiation fields. In Section~\ref{sec:obs} we confront our numerical calculations with the IceCube result and put constraints on the properties of such prospective neutrino sources.

\section{Neutrinos from AGN's with Shakura-Sunyaev accretion disk}
\label{sec:SS}

AGN's are long sought potential sites for high energy neutrino production. They can accelerate protons up to highest energies and they are surrounded by high intensity radiation fields were photo-nuclear reactions with subsequent neutrino emission can occur.
At the heart of an AGN resides super-massive black hole surrounded by the accretion disc.
Accretion disc is hot and is emitting thermal radiation which gives prominent feature in the observed AGN spectra usually refereed to as a "Big Blue Bump". Accelerated particles move along two jets perpendicular to the accretion disc and crossing this radiation field.

In what follows we employ the following model for neutrino production. We assume that proton acceleration occurs directly near the black hole horizon, see e.g. Refs.~\cite{Neronov:2007mh,Kalashev:2012cm}. High energy neutrino appear in charged pion decays created in $p\gamma \rightarrow n\pi^+$ and $n\gamma \rightarrow p\pi^-$ reactions in collisions with "blue bump" photons. As a first step we remind observational phenomenology of accretion discs and estimate optical depth for these photopion production reactions.

\subsubsection{Accretion discs phenomenology}

The effective temperature of optically thick material on the scale of gravitational radius is given by~\cite{Shakura}
\begin{equation}
T_0 = 30 ~{\rm eV}~\left(\frac{M}{10^8 M_\odot}\right)^{-1/4} \left(\frac{L}{\eta\, L_{\rm Edd}}\right)^{1/4},
\label{eq:T_0}
\end{equation}
where $M$ is  mass of a black hole and $\eta$ is the efficiency of converting gravitational potential energy to electromagnetic radiation, $L = \eta\dot{M}$, at given accretion rate $\dot{M}$. Eddington luminosity, $L_{\rm Edd}$, is defined as
$$ L_{\rm Edd} = 1.26 \cdot 10^{46} \left(\frac{M}{10^8 M_\odot}\right) {\rm erg ~s}^{-1}.$$

Temperature has power low profile with radial coordinate on the disc, $T \propto r^{-\beta}$.
In theory~\cite{Shakura} $\beta = 3/4$. Observationally, the slope is consistent with thin disc theory, $\beta = 0.61^{+0.21}_{-0.17}$, but would also allow a shallower temperature profile that would reduce the differences between the microlensing and flux size estimates~\cite{Morgan:2007im}.

Within uncertainties and with accuracy sufficient for our purposes, the observed disc sizes at radiation frequency $E_\gamma = 5$~eV can be fitted  by the relation~\cite{Morgan:2010xf}
$$R = 10^{15} ~{\rm cm}~ \left(\frac{M}{10^8 M_\odot}\right),$$
which is about two orders of magnitude larger as compared to the gravitational radius.
This estimate is somewhat larger as compared to the expectation from thin disk theory.
The photon density around the disc can be approximated by the relation
$$ n_\gamma = \frac{L_{\rm disk}}{4\pi R^2 E_c}, $$
where $E_c$ is typical photon energy. On average, Spectral Energy Distributions (SED) of AGNs are peaked at energy $E_c = 10$ eV, for a review see e.g. Ref.~\cite{Koratkar_Blaes}.

The optical depth to photomeson production can be estimated as $\tau = \sigma n_\gamma R$, where $\sigma \approx 5\times 10^{-28}~{\rm cm}^2$ is cross section at $\Delta$-resonance. This gives 
$$ \tau \sim 10^3 \; \left(\frac{L_{\rm disk}}{L_{\rm Edd}}\right)\; \left(\frac{10~{\rm eV}}{E_c}\right),$$
irrespective of the black hole mass. 
There are tight correlations between monochromatic and bolometric luminosities of AGN, e.g. $\lambda L_{\lambda}(5100\; A^\circ) \approx 0.1\, L_{\rm bol}$, see  \cite{Elvis:1994us,Richards:2006xe}. $\lambda L_{\lambda}$ gives estimate for $L_{\rm disk}$. For typical bolometric luminosity we can assume $L_{\rm bol} \approx 0.1\, L_{\rm Edd}$, see e.g. Ref.~\cite{Woo:2002un,Kaspi:1999pz}.  Therefore $\tau \sim 10$ would be typical value for the optical depth to photomeson production after traveling distance comparable to the accretion disc size. 


\subsubsection{Radiation fields and reaction rates.}
\label{model}

In the laboratory frame, the rate of reactions with the photon
background is given by the standard expression,
\begin{equation}
R = \int d^3p\, n({\bf p}) (1-\cos\theta) 
\sigma(\tilde\omega),
\label{eq:rate-generic}
\end{equation}
where $n({\bf p})$ is the photon density in the laboratory frame,
$\sigma(\tilde\omega)$ is the cross section of the relevant reaction
in the rest frame of the primary particle as the function of the
energy of the incident photon $\tilde\omega=\gamma p(1 -
\cos\theta)$, $\gamma$ is the gamma-factor of the primary particle in
the laboratory frame.

For the  black body radiation with
temperature $T$ one has
\begin{equation}
n_{\rm T}(p) \equiv {2\over (2\pi)^3} {1\over \exp(p/T) -1}.
\label{eq:n_bb}
\end{equation}
We assume that the disc segment at a radius $r$ radiates black body with local temperature $T(r)$, 
\begin{equation}
T(r) = T_0 \,F(r),
\end{equation}
where $T_0$ is given by Eq.~(\ref{eq:T_0}), and $F(r)$ is given by~\cite{Shakura}
\begin{equation}
F(r) = \left(\frac{r_g}{r}\right)^{3/4}\left(1-\sqrt{\frac{r_{in}}{r}}\,\right)^{1/4}.
\label{Eq:Tprofile}
\end{equation}
Here $r_g=2\kappa M$ is Schwarzschild gravitational radius, 
$$
r_g = 3\times 10^{13}\; \left(\frac{M}{10^8 M_\odot}\right) {\rm cm},
$$
and $r_{in}^{~}$ is the radius of the disk inner edge. 
Contribution of such segment to the photon density at point $z$ along disc axis is
\begin{equation}
n({\bf p})  = \frac{\delta({\bf n-n}_0)\; rdr}{(r^2+z^2)} \;\; n_{\rm T}(p,r),
\label{eq:eps_gal}
\end{equation}
where  ${\bf n}_0$ is the unit
vector in the direction from $r$ to $z$. Its contribution to 
the reaction rate Eq.~(\ref{eq:rate-generic}) can
be expressed as
\begin{equation}
R(z,r,\gamma) =
\frac{(1-\cos\theta)}{4\pi^3(r^2+z^2)}\int_0^\infty \frac{dp\,
p^2\, \sigma(\tilde\omega)}{e^{p/T(r)}-1},
\label{eq:rate-point}
\end{equation}
where 
$\cos\theta = {z}/{\sqrt{r^2+z^2}}.$
Finally
\begin{equation}
R(z,\gamma) = 2\pi \int_{r_{in}^{~}}^\infty r dr R(z,r,\gamma),\label{eq:rate}
\end{equation}

The disk inner edge $r_{in}^{~}$ is related to the radiation efficiency as 
$$\eta = \frac{3}{2}\int_{r_{in}^{~}}^\infty rdrF^4(r).$$
In what follows we use $\eta = 0.1$, which is usual assumption in existing literature.

\begin{figure}
\includegraphics[width=0.5\textwidth]{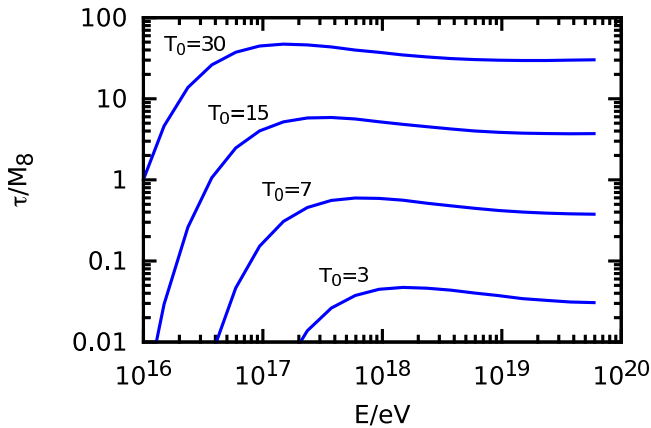}
\caption{Optical depth as function of proton energy for several values of $T_0$ (in eV).}
\label{no_escape}
\end{figure}

Optical depth with respect to this reaction for protons accelerated near black hole horizon and moving along jet axis from $z_0$ to infinity is given by $\tau(\gamma) = \int_{z_0}^\infty dz R(z,\gamma)$.
Resulting function $\tau(E)$ for photomeson production is shown in Fig. 1 for several values of $T_0$~\footnote{Note that the average background photon energy  is roughly $0.1 T_0$ assuming temperature profile~(\ref{Eq:Tprofile}) and radiation efficiency $\eta=0.1$} and $z_0 = r_g$. To produce neutrinos with energy $E_\nu \sim 10^{15}~\rm eV$ efficiently, one needs the optical depth with respect to this reaction to be larger than unity for protons with  $E \sim 10^{17}~\rm eV$. This requirement translates to $T_0 > 10~ \rm eV$,  see~Fig.~\ref{no_escape}.

\section{The observed spectrum}
\label{sec:obs}

In this paper we do not study the processes of particle acceleration but simply assume that protons are accelerated by electric fields in close vicinity of the black hole horizon, for relevant models see e.g. Ref.~\cite{Neronov:2007mh}. For definiteness we assume that the  spectrum of accelerated protons at  $z=z_0$ has the power-law form
\begin{equation}
j_{acc}(E)=E^{-\alpha}, \;  E<E_{max},
\label{primarySpec}
\end{equation}
their momenta are directed along the disk axes, and at larger $z$ electric fields and acceleration proceses are negligible. In what follows we choose $\alpha = 2$ and $z_0 = 2r_g$. 

The calculation of the observable neutrino spectrum is performed in two steps. First we simulate propagation of protons through the radiation field at $z > z_0$ and  calculate resulting spectrum of nucleons and of products of their interaction. To be conservative we assume that the magnetic field is negligible and therefore both protons and neutrons with energies below the pion production threshold may freely escape the source. This will let us calculate the maximal possible contribution of the process to the observed spectrum of cosmic rays. As for the secondary electron-photon cascade calculation, to obtain it's upper bound we will always assume that the cascade freely escapes the source region and show below that even in this extreme case the predicted contribution of this process to the diffuse photon background would be far below the present observational upper limits.

We model interactions employing Monte-Carlo approach. In particular,  during the $i$-th iteration at position $z=z_i$ the traveled optical depth $\tau_i$ is sampled using the equation
$$
\tau_i = -log(\xi).
$$
Here and below  $\xi$ is uniformly distributed random variable, $0<\xi<1$. The point of next interaction $z_{i+1}$ is calculated by solving equation
\begin{equation}
\int_{z_i}^{z_{i+1}}R(z,\gamma_i)dz = \tau_i \label{eq:iteration},
\end{equation}
where $\gamma_i$ is current gamma factor of the nucleon and $R(z,\gamma_i)$ is interaction rate given by Eq.(\ref{eq:rate}). The background photon momentum is sampled in each interaction point $z_i$ in the following way. Firstly disk segment $r_i$ emitting the photon is sampled using Eq.~(\ref{eq:rate}):
\begin{equation}
\xi R(z,\gamma) = 2\pi \int_{r_{in}^{~}}^{r_i} r dr R(z_i,r,\gamma),
\end{equation}
then photon energy $p_i$ is sampled using Eq.(~\ref{eq:rate-point}):
\begin{equation}
\xi R(z,r,\gamma) =
\frac{(1-\cos\theta_i)}{4\pi^3(r_i^2+z_i^2)}\int_0^{p_i} \frac{dp\,
p^2\, \sigma(\tilde\omega)}{e^{p/T(r_i)}-1},
\end{equation}
where $\cos\theta_i=z_i/\sqrt{z_i^2+r_i^2}$. Finally SOPHIA event generator~\cite{Mucke:1999yb} is used to sample the recoiling nucleon energy $\gamma_{i+1}$ as well as secondary particles and their momenta. The iterations continue while Eq~(\ref{eq:iteration}) has solution.
Absence of solution means that the nucleon freely escapes the AGN site. As a result we obtain the spectrum of nucleons, neutrinos, photons and electrons leaving the interaction region.

In the second step we integrate over the distribution of sources taking into account their possible abundance evolution and cosmological propagation effects~\footnote{We do not make averaging over $T_0$ and $E_{\rm max}$.}. The latter mostly reduces to red shift, neutron decay, and electron-photon cascade development lead by interactions with CMB and intergalactic infrared background. This procedure is performed with the use of the numerical code developed in Ref.~\cite{propagation}. The code simulates interactions of nucleons, photons and stable leptons with intergalactic photon backgrounds. For nucleons it takes into account photopion production, $e^+e^-$-pair production and neutron decay. The secondary particles produced in these interactions are also traced in the code. The electron-photon cascade is mostly driven by the chain of inverse Compton scattering of electrons on background photons and $e^+e^-$-pair production by photons. We also take into account neutrino mixing using the mixing angles in the tribimaximal approximation,  which is sufficient with current limited statistics. The resulting spectrum has a flavor ratio of approximately (1:1:1).

Finally, we normalize the simulated spectra using IceCube data. Namely, using 22 events with deposited energy above $42~\rm TeV$, published in Ref.~\cite{Icecube3years}, and exposure dependence on energy from Ref.~\cite{Icecube2years}, we maximize Poisson probability of observing the above events provided that a given theoretical model is true.

\begin{figure}
\includegraphics[height=0.7\linewidth,angle=0]{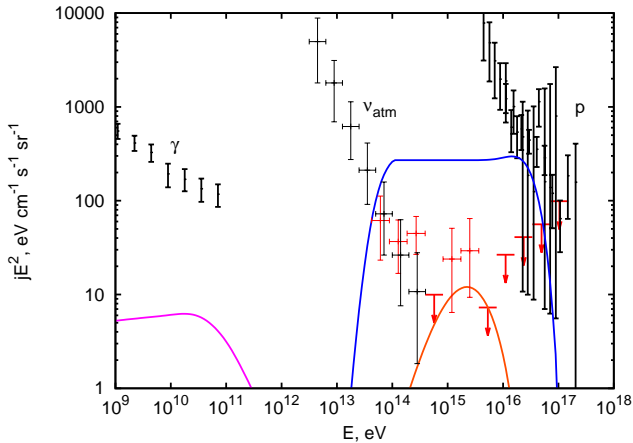}
\caption{Secondary neutrino (red line) and gamma-ray (magenta line) fluxes from protons (blue line) with $1/E^2$ power law injection spectrum and $E_{max}=100$ PeV for disk temperature $T_0=15$ eV, black hole mass $10^8 M_\odot$ and luminosity evolution of sources  $(1+z)^3$. Red points with errorbars represent the IceCube astrophysical neutrino flux from Ref.\cite{Icecube3years}. Atmospheric neutrino flux is taken from Ref.\cite{IceCube_atmosphere}, Fermi diffuse gamma-ray flux is from Ref.~\cite{Fermi_EGB}, while proton flux is from Ref.~\cite{dataKG}.}
\label{example1}
\end{figure}

\begin{figure}
\includegraphics[height=0.7\linewidth,angle=0]{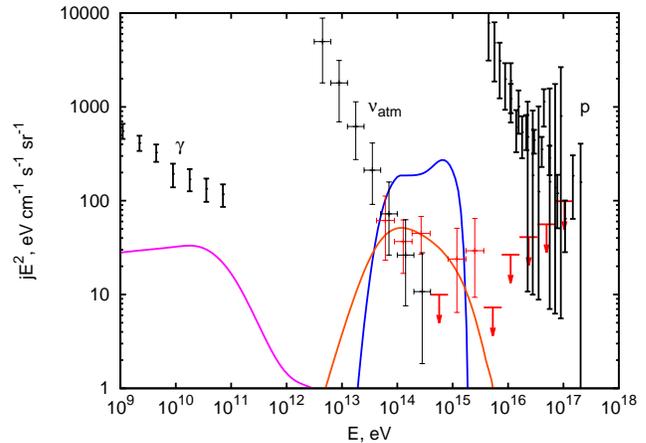}
\caption{Secondary neutrino and gamma-ray fluxes from protons with $1/E^2$ power law spectrum and $E_{max}=30$ PeV for $T_0=120$ eV and luminosity evolution of sources  proportional to $(1+z)^3$. Experimental data are the same as in Fig.~\ref{example1}.}
\label{example2}
\end{figure}

In the energy bin 0.4-1 PeV the IceCube does not have any events in present data. 
According to Ref.~\cite{Icecube3years} a gap larger than this one appears in 43\% of realizations of the best fit continuous spectra. Therefore, one may safely assume that the real neutrino spectrum is a smooth power law. On the other hand, different energy regions may correspond to different populations of sources, and, therefore, spectrum may have features. E.g. peak at $E_\nu \sim 2~\rm PeV$ might be real. At present one should consider both possibilities and we follow this line of thought in presenting results.

In the  Fig.~\ref{example1} we present secondary neutrino  flux (shown in red) from protons accelerated to $E_{max}=100$ PeV   and absorbed in the disk radiation field with  temperature $T_0=15$ eV (black hole mass $10^8 M_\odot$ and luminosity evolution of sources  $(1+z)^3$ is assumed). We see that resulting neutrino spectrum is rather narrow~\footnote{We have studied the models with monochromatic injection spectra of accelerated protons as well. Resulting neutrino spectra are somewhat narrower, as compared to a power law injection, but are not monochromatic.}  and therefore population of objects with such low temperature may explain narrow bumps in the spectrum. 

Fig.~\ref{example2} corresponds to $T_0=120$ eV and $E_{max}=30$ PeV. In this case all high energy part of the IceCube neutrino flux  at $E>100$ TeV can be explained, assuming that the absence of events in the energy bin  0.5-1 PeV is due to a statistical fluctuation.  One would have to explain low energy data $E<100$ TeV with other type of sources still, if such data will appear, since our model has low energy cutoff at 100 TeV due to the energy threshold for photopion production.

Calculated secondary photon flux in all cases is significantly below diffuse $\gamma$-background, measured by Fermi. Measured~\cite{dataKG} proton flux at energies $E=1-100$ EeV is dominated by Galactic sources, see for example, Ref.~\cite{escape_knee}. Therefore, contribution of extragalactic sources to the observed proton flux should be sub-dominant. Our results do not contradict to this observation as well.

\begin{figure}[t]
\includegraphics[height=0.7\linewidth,angle=0]{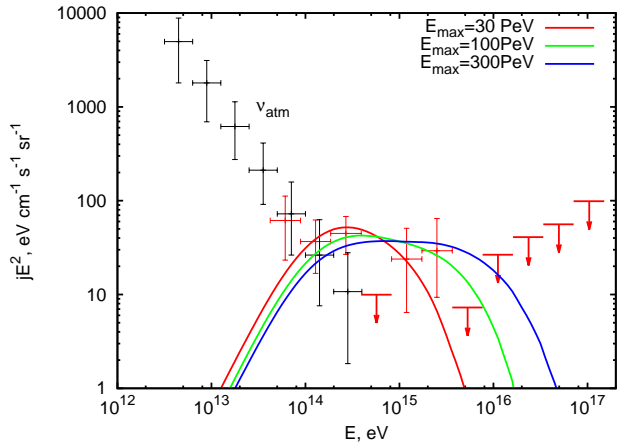}
\caption{Dependence of neutrino flux on maximum proton energy $E_{max}$ for $E^{-2}$ power law injection spectrum, disk temperature $T_0=60$ eV and luminosity evolution of sources  $\propto (1+z)^3$. Red points with errorbars show the IceCube astrophysical neutrino flux after 3 years of exposure, taken from Ref.\cite{Icecube3years}. Atmospheric neutrino flux, Ref.~\cite{IceCube_atmosphere}, is shown by black points with errorbars.}
\label{Emax}
\end{figure}

\begin{figure}[t]
\includegraphics[height=0.7\linewidth,angle=0]{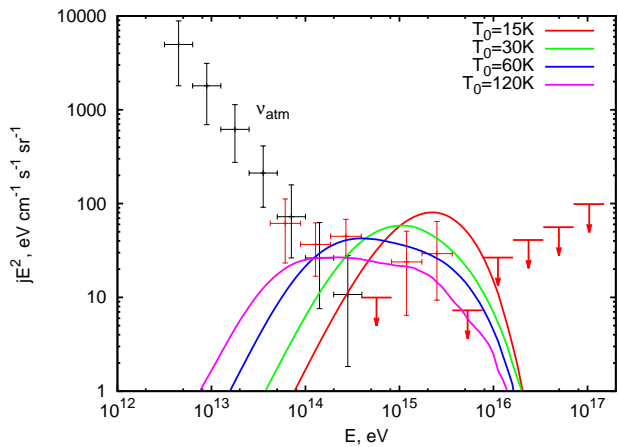}
\caption{Dependence of neutrino flux on disk temperature $T_0$ for $E_{max}=100$ PeV. The remaining parameters and  experimental data are the same as in Fig.~\ref{Emax}.}
\label{Temp}
\end{figure}

Model dependence of resulting neutrino flux is shown in Figs.~\ref{Emax}-\ref{Temp}.
In the Fig.~\ref{Emax} the dependence on the maximum energy of accelerated protons is presented~for $T_0=60$ eV.  Models with maximum energies $E_{\rm max} = 30$ PeV, $E_{\rm max} = 100$ PeV and $E_{\rm max} = 300$ PeV are shown by red, green and blue lines correspondingly. 
In the Fig.~\ref{Temp} we present dependence of the neutrino flux on disk temperature for $E_{\rm max} = 100$ PeV.  Disk temperatures $T_0=15$ eV, $T_0=30$ eV, $T_0=60$ eV and $T_0=120$ eV  are shown by red, green, blue and pink lines.  The spectrum of neutrinos for $T_0=15$ eV is peaked at 1-3 PeV and may be responsible for the high energy part of the IceCube data. The case of high temperature $T_0=120$ eV can explain the IceCube data for the whole energy range $E>100$ TeV.


\section{Discussion}

In this paper we have made an attempt to explain the extragalactic neutrino signal recently announced by the IceCube collaboration~\cite{Icecube3years}. As prospective class of neutrino sources we have chosen AGNs. To model neutrino creation we study photopion production processes on the radiation field of the Shakura-Sunyaev accretion disks in the black hole vicinity. To our knowledge our work is the first one where the realistic anisotropic radiation field of the accretion disc was considered for these purposes.

Important parameters describing the model are maximum energy of accelerated protons and disk temperature. We have studied the parameter space of the model and compared  predicted neutrino fluxes with the IceCube measurement. Along the way we took into account constraints set by the diffuse gamma-ray background measurements by the Fermi observatory~\cite{Fermi_EGB} and by the proton flux measurements by KASCADE and KASCADe-Grande experiments~\cite{dataKG}. We have shown that the model presented in this paper can naturally explain the neutrino  spectrum observed by the IceCube. The model can be falsified (and better constrained) studying correlation signal between neutrino arrival directions and varios sub-classes of AGNs. Such study will became feasible in the nearest future, as more data will accumulate. Current strong directional limits on possible neutrino sources are given, e. g. in Ref.~\cite{Adrian-Martinez:2014wzf,Aartsen:2014cva}.

\begin{acknowledgments}
This work was supported by the Russian Science Foundation, grant 14-12-01340. Numerical calculations have been performed at the computer cluster of the Theoretical Physics
Division of the Institute for Nuclear Research of the Russian Academy of Sciences.
\end{acknowledgments}


\end{document}